\documentclass[11pt]{article}
\usepackage{graphicx,slashed,multirow}
\usepackage[margin=1.25in]{geometry}
\usepackage[usenames,dvipsnames]{color}
\usepackage{xcolor}
\usepackage{url}
\usepackage[colorlinks = true,
            linkcolor = blue,
            urlcolor  = blue,
            citecolor = blue,
            anchorcolor = blue]{hyperref}
\usepackage{amsmath}
\usepackage{subcaption}
\usepackage{cite}
 \usepackage[normalem]{ulem}

\usepackage[capitalise]{cleveref}
\crefname{section}{Sec.}{Secs.}
\crefname{table}{Tab.}{Tabs.}
\crefname{figure}{Fig.}{Figs.}
\crefname{equation}{Eq.}{Eqs.}
\crefname{appendix}{Appendix}{Appendix}

\usepackage{booktabs}
\newcommand{\ra}[1]{\renewcommand{\arraystretch}{#1}}

\newcommand{\fsmog}{\texttt{FSMOG}~}

\newcommand\pubnumber{KIAS-A22001}
\newcommand\pubdate{\today}

\def\Title#1{\begin{center} {\LARGE #1 } \end{center}}
\def\Author#1{\begin{center}{ \sc #1} \end{center}}
\def\Address#1{\begin{center}{ \it #1} \end{center}}

\newcommand\pubblock{\rightline{\begin{tabular}{l} \pubnumber\\
         \pubdate \end{tabular}}}
\newenvironment{Abstract}{\begin{quotation} \begin{center}
                       ABSTRACT
     \end{center}\bigskip  }{\end{quotation}}

\def\beq{\begin{equation}}
\def\eeq#1{\label{#1}\end{equation}}
\def\eeqn{\end{equation}}

\newenvironment{Eqnarray}%
   {\arraycolsep 0.14em\begin{eqnarray}}{\end{eqnarray}}
\def\beqa{\begin{Eqnarray}}
\def\eeqa#1{\label{#1}\end{Eqnarray}}
\def\eeqan{\end{Eqnarray}}

\let\bar=\overbar

\def\lsim{\mathrel{\raise.3ex\hbox{$<$\kern-.75em\lower1ex\hbox{$\sim$}}}}
\def\gsim{\mathrel{\raise.3ex\hbox{$>$\kern-.75em\lower1ex\hbox{$\sim$}}}}

\def\del{\partial}
\def\Dslash{\not{\hbox{\kern-4pt $D$}}}
\def\dslash{\not{\hbox{\kern-2pt $\del$}}}
\def\pslash{\not{\hbox{\kern-2pt $p$}}}
\def\ETmiss{\not{\hbox{\kern-4pt $E$}}_T}

\def\Dlr{\mathrel{\raise1.5ex\hbox{$\leftrightarrow$\kern-1em\lower1.5ex\hbox{$D$}}}}

\def\MSB{{\bar{M \kern -2pt S}}}
\def\msb{{\bar{\scriptsize M \kern -1pt S}}}

\def\drb{{\bar{\scriptsize D \kern -1pt R}}}

\def\TeV{{\rm TeV}}

\newcommand\snowmass{\begin{center}\rule[-0.2in]{\hsize}{0.01in}\\\rule{\hsize}{0.01in}\\
\vskip 0.1in Submitted to the  Proceedings of the US Community Study\\ 
on the Future of Particle Physics (Snowmass 2021)\\ 
\rule{\hsize}{0.01in}\\\rule[+0.2in]{\hsize}{0.01in} \end{center}}

\begin{document}

\pubblock

\Title{Phenomenological aspects of composite Higgs scenarios:\\exotic scalars and vector-like quarks}

\Author{
A.~Banerjee$^1$,
D.~Buarque Franzosi$^2$,
G.~Cacciapaglia$^3$,
A.~Deandrea$^3$,
G.~Ferretti$^1$,
T.~Flacke$^4$,
B.~Fuks$^5$,
M.~Kunkel$^6$,
L.~Panizzi$^7$,
W.~Porod$^6$, 
L.~Schwarze$^6$
}


\Address{$^1$ Department of Physics, Chalmers University of Technology, Fysikg\aa rden, 41296 G\"oteborg, Sweden}
\vspace{-9mm}
\Address{$^2$ Stockholm University, Department of Physics,
106 91 Stockholm, Sweden}
\vspace{-9mm}
\Address{$^3$ Univ. Lyon, Universit{\' e} Claude Bernard Lyon 1, CNRS/IN2P3,
 IP2I UMR5822, F-69622, Villeurbanne, France}
\vspace{-9mm}
\Address{$^4$ Center for AI and Natural Sciences, KIAS, Seoul 02455, Korea}
\vspace{-9mm}
\Address{$^5$ Laboratoire de Physique Th\'eorique et Hautes Energies (LPTHE),
  UMR 7589, Sorbonne Universit\'e et CNRS, 4 place Jussieu, 75252 Paris Cedex 05, France}
\vspace{-9mm}
\Address{$^6$ Institut f\"ur Theoretische Physik und Astrophysik,  Uni W\"urzburg, Emil-Hilb-Weg 22,
D-97074 W\"urzburg, Germany}
\vspace{-9mm}
\Address{$^7$ Department of Physics and Astronomy, Uppsala University, Box 516, SE-751 20 Uppsala, Sweden}

\vspace{-12mm}
\snowmass
\vspace{-12mm}

\begin{Abstract}
\noindent
Composite Higgs models usually contain additional pseudo Nambu Goldstone bosons and vector-like quarks. We discuss various aspects related to their LHC phenomenology and provide summary plots of exclusion limits using currently available information. We also describe a general parametrisation implemented in a software for Monte Carlo simulations and study the $SU(5)/SO(5)$ scenario as a concrete example.
\end{Abstract}

\def\thefootnote{\fnsymbol{footnote}}
\setcounter{footnote}{0}
\newpage
\tableofcontents
\bigskip 
\hrule 
\addtocontents{toc}{\protect\setcounter{tocdepth}{1}}

\section{Executive summary}
Models of electroweak (EW) symmetry breaking by a composite Higgs~\cite{Kaplan:1983fs} with partial compositeness~\cite{Kaplan:1991dc} in the top-quark sector give promising solutions to the hierarchy problem~\cite{Agashe:2004rs,Contino:2010rs,Bellazzini:2014yua,Panico:2015jxa,Cacciapaglia:2020kgq} of the Standard Model (SM). At the effective field theory level they can be described by specifying the pattern of symmetry breaking involved. Generically these models predict the existence of light scalars in addition to the Higgs boson, all emerging as pseudo Nambu-Goldstone bosons (pNGBs). Partial compositeness also implies the presence of vector-like quarks (VLQs). This proliferation of new particles requires a systematic approach to study this class of models and to facilitate a seamless transition between theory, data, and simulation codes. 

Current experimental analyses already allow to set limits on the masses of such particles in specific scenarios depending on their interactions. For pNGBs, we provide an overview plot of the limits for pair production cross-sections times branching ratios in various final states (\cref{fig:dyquarkscombined}). We also work out a concrete example based on the $SU(5)/SO(5)$ coset (\cref{fig:splittingplotb}). For VLQs, we emphasize the relevance of their exotic decays by presenting a plot summarizing current limits in the VLQ mass versus pNGB mass plane (\cref{fig:VLQSummaryPlot}). 

Furthermore, we provide a general parametrisation describing the interactions of the new particles arising in composite scenarios, which can be used for phenomenological studies and is implemented in a numerical model for Monte Carlo simulations\cite{VLQpNGB}. A software toolbox of a new kind is also being developed, \fsmog (\texttt{FeynRules} plugin for Simplified MOdel Generation) \cite{fsmog}, to automatize the construction of the most general interaction Lagrangian given a specific particle content, which has a much wider range of applications beyond the realm of composite Higgs models.

\section{Composite Higgs and partial compositeness: an outline} \label{sec:theoryoverview}

There are many approaches to strongly-coupled EW symmetry breaking scenarios~\cite{Contino:2010rs,Bellazzini:2014yua,Panico:2015jxa,Cacciapaglia:2020kgq}. The most investigated ones include, e.g., 5D holographic theories~\cite{Contino:2003ve} and multi-site deconstructed models like Little Higgs models  \cite{Schmaltz:2005ky}. Instead of using the holographic duality, we consider directly a class of composite Higgs models based on 4D asymptotically free (hypercolor) gauge theories. We assume that the hypercolor theory, after going through a near conformal running \cite{Holdom:1981rm,Cohen:1988sq}, confines at the multi-TeV scale. 
To include top partial compositeness, its fundamental degrees of freedom contain fermionic matter in two inequivalent irreducible representations (irreps) of hypercolor~\cite{Barnard:2013zea,Ferretti:2013kya}, chosen in order to sequester the EW coset from the composite states carrying QCD color.

The strong sector is associated with a global symmetry $\mathcal{G}$ that breaks spontaneously to a subgroup $\mathcal{H}$ below the confinement scale, such that the SM gauge symmetry is contained in the unbroken subgroup $\mathcal{H}$. The symmetry breaking pattern induced by the fermionic condensates gives rise to an EW coset containing the Higgs boson as a pNGB. The three minimal cosets arising from real, pseudoreal and complex irreps, containing at least a pNGB Higgs doublet and preserving custodial symmetry, are $SU(5)/SO(5)$ \cite{Dugan:1984hq,Ferretti:2014qta} (real), $SU(4)/Sp(4)$ \cite{Galloway:2010bp,Cacciapaglia:2014uja} (pseudoreal), and $SU(4) \times SU(4)^\prime / SU(4)_D$ \cite{Ma:2015gra,Wu:2017iji} (complex), respectively. 
Holographic/little Higgs constructions, reviewed in \cite{Contino:2010rs} and \cite{Schmaltz:2005ky}, lead to additional cosets, for instance the well-known minimal model $SO(5)/SO(4)$ (MCHM) \cite{Agashe:2004rs} or $SO(6)/SO(4)\times SO(2)$ \cite{Mrazek:2011iu}. Most of the low energy phenomenology discussed here is also applicable to this wider variety of models. Below the condensation scale, the colored VLQs originate from the trilinear fermionic operators combining the two irreps in a hypercolor invariant way. They are assumed to mix linearly with the SM quarks of the third generation through partial compositeness.
We do not review these constructions here but refer to the existing literature: see \cite{Barnard:2013zea, Ferretti:2013kya} for the initial constructions (a streamlined classification can be found in \cite{Ferretti:2016upr}). Additional models of this type are presented in \cite{Ferretti:2014qta,Vecchi:2015fma,Elander:2020nyd,Erdmenger:2020lvq,Erdmenger:2020flu,Elander:2021bmt,Gherghetta:2020ofz,Appelquist:2020bqj,Cacciapaglia:2019dsq,Bizot:2016zyu}.

Composite Higgs models have two major types of experimental signatures in the low energy range. First, the pNGB nature of the Higgs boson implies modifications of its couplings with the other SM particles \cite{Montull:2013mla,Sanz:2017tco,Liu:2017dsz,Banerjee:2017wmg,Liu:2018qtb}. The other kind of signature involves additional particles beyond the SM (BSM), predicted by these models, for example, exotic EW pNGBs \cite{Arbey:2015exa,Ferretti:2016upr,Belyaev:2016ftv,Agugliaro:2018vsu}, colored pNGBs \cite{Cacciapaglia:2015eqa,Belyaev:2016ftv}, VLQs \cite{Bizot:2018tds,Cacciapaglia:2019zmj,Xie:2019gya,Matsedonskyi:2014lla,Corcella:2021mdl,Dasgupta:2021fzw,Chala:2017xgc} and other colored fermionic states in non-triplet irreps \cite{Cacciapaglia:2021uqh}, vector resonances \cite{Azatov:2015xqa,BuarqueFranzosi:2016ooy,Yepes:2018dlw,Dasgupta:2019yjm,Banerjee:2021efl}, and axion-like particles \cite{Cacciapaglia:2019bqz,BuarqueFranzosi:2021kky}. In this contribution we focus on EW pNGBs and exotic decay modes of VLQs.

To construct the low energy theory~\cite{Coleman:1969sm,Callan:1969sn} involving the pNGBs of the EW sector, the VLQs, and the SM quarks and gauge bosons one must choose a symmetry breaking pattern $\mathcal{G}/\mathcal{H}$, VLQs in $\mathcal{H}$ irreps, and an embedding of the SM quarks into an irrep of $\mathcal{G}$, see  \cite{DeSimone:2012fs,Marzocca:2012zn,Panico:2015jxa,Bizot:2018tds, Banerjee:2022izw} for detailed constructions. 

\subsection{A general parametrisation}
\label{sec:fsmog}

Here and in the appendix we present the generic Lagrangian, focusing on the interactions of the pNGBs and the VLQs. At this stage we only use invariance under the electromagnetic $U(1)_{\textrm{em}}$ and the QCD color $SU(3)_c$ to construct the interactions. This introduces a large number of arbitrary couplings that will have to be fixed in a specific model implementation. We take this approach since it can be easily implemented in software tools.

We consider the particle content shown in  \cref{tab:particles}. As mentioned above, even this choice reflects some restrictions on the type of degrees of freedom present. However, this field content is sufficient for the discussion below.
\begin{table}[t!]
	\def\arraystretch{1.3}
	\centering
	\begin{tabular}{ccccccccc}
		\toprule 
		Fields & Spin & $SU(3)_c$ & $U(1)_{\textrm{em}}$\\  
		\midrule 
		$S^0_i$ & $0$ & $\bf 1$ & 0 \\
		$S^\pm_i$ & $0$ & $\bf 1$ & $\pm 1$ \\
		$S^{\pm\pm}_i$ & $0$ & $\bf 1$ & $\pm 2$ \\
		\midrule 
		$\pi^{q}_r$ & $0$ & $\mathbf{r}$ & $q$ \\
		\midrule 
		$T$ & $1/2$ & $\bf 3$ & $2/3$ \\
		$B$ & $1/2$ & $\bf 3$ & $-1/3$ \\
		$X$ & $1/2$ & $\bf 3$ & $5/3$ \\
		\bottomrule 
	\end{tabular} 
	\caption{ BSM particle content considered in the Lagrangian presented in this section. For the colored scalars $\pi^{q}_r$, the subscript $r=\mathbf{3}, \mathbf{6}, \mathbf{8}$ denotes the $SU(3)_c$ representation. We allow for multiple colorless scalars $S^{0, \pm, \pm\pm}_i$ $(i=1,2\dots)$ since they generically arise in composite Higgs models.}
	\label{tab:particles}
\end{table}
The interaction Lagrangian can be expressed as follows
\begin{align}
\mathcal{L}_{\rm int}&=\mathcal{L}_{SS V}+\mathcal{L}_{SS VV}+\mathcal{L}_{S VV}
+\mathcal{L}_{S V\tilde{V}}
+\mathcal{L}_{\Psi \Psi V}+\mathcal{L}_{\Psi f V}+\mathcal{L}_{\Psi f S}+\mathcal{L}_{f f S}\nonumber \\
& + \mathcal{L}_{\pi\pi V+\pi\pi VV}
+\mathcal{L}_{\pi V\tilde{V}}
+ \mathcal{L}_{\Psi f \pi}+\mathcal{L}_{f f \pi}\,,
\label{L_int}
\end{align}
where $\Psi\equiv\{T,B,X\}$ collectively denotes the VLQs. In addition to the field content in \cref{tab:particles}, $V$ and $f$ denote the SM gauge bosons and third generation quarks. The explicit expressions for each term of the above Lagrangian are presented in the appendix.

For the time being, we use a \texttt{FeynRules} implementation of the above Lagrangian, limited to the EW pNGBs, and VLQs \cite{VLQpNGB}, through which a UFO model \cite{Degrande:2011ua} has been generated. The UFO implementation allows the simulation of processes up to one-loop in QCD. 
A more automatized tool, \fsmog (\texttt{FeynRules} plugin for Simplified MOdel Generation), is being developed to facilitate the writing of the most general Lagrangian. The program will take as input fields of given spin, charge, and color and output the \texttt{FeynRules} file containing the most general interactions between these fields and the SM ones, up to a certain mass dimension. At present this is limited to dimension-4 operators for simplicity, however future versions of \fsmog will include commonly used higher dimension operators. \fsmog is still under development, but the current version can be obtained from the authors \cite{fsmog}. In this implementation the reduction of number of independent couplings for a specific model is achieved by using a \texttt{FeynRules}-style restriction card. 

\section{LHC bounds for pair production of pNGBs and VLQs}
\label{sec:colliderpheno}

\subsection{Electroweak pNGBs}

In this section we confine ourselves to the case where the only accessible degrees of freedom are the pNGBs from the EW coset. We present the bounds on their production cross sections times branching ratios as function of their mass in various channels. In this way, the bounds are applicable to any BSM scenario with scalars.

\begin{figure}
    \centering
    \includegraphics{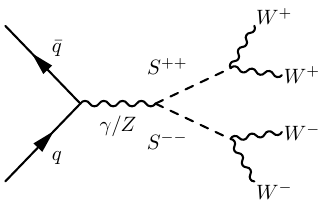} \quad \includegraphics{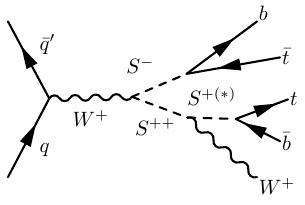} 
    \caption{Examples of pNGB pair production via Drell-Yan processes with subsequent decays into SM particles.}
    \label{fig:pNGBpairgraphs}
\end{figure}

\begin{figure}[th]
	\centering
	\includegraphics[width=\textwidth]{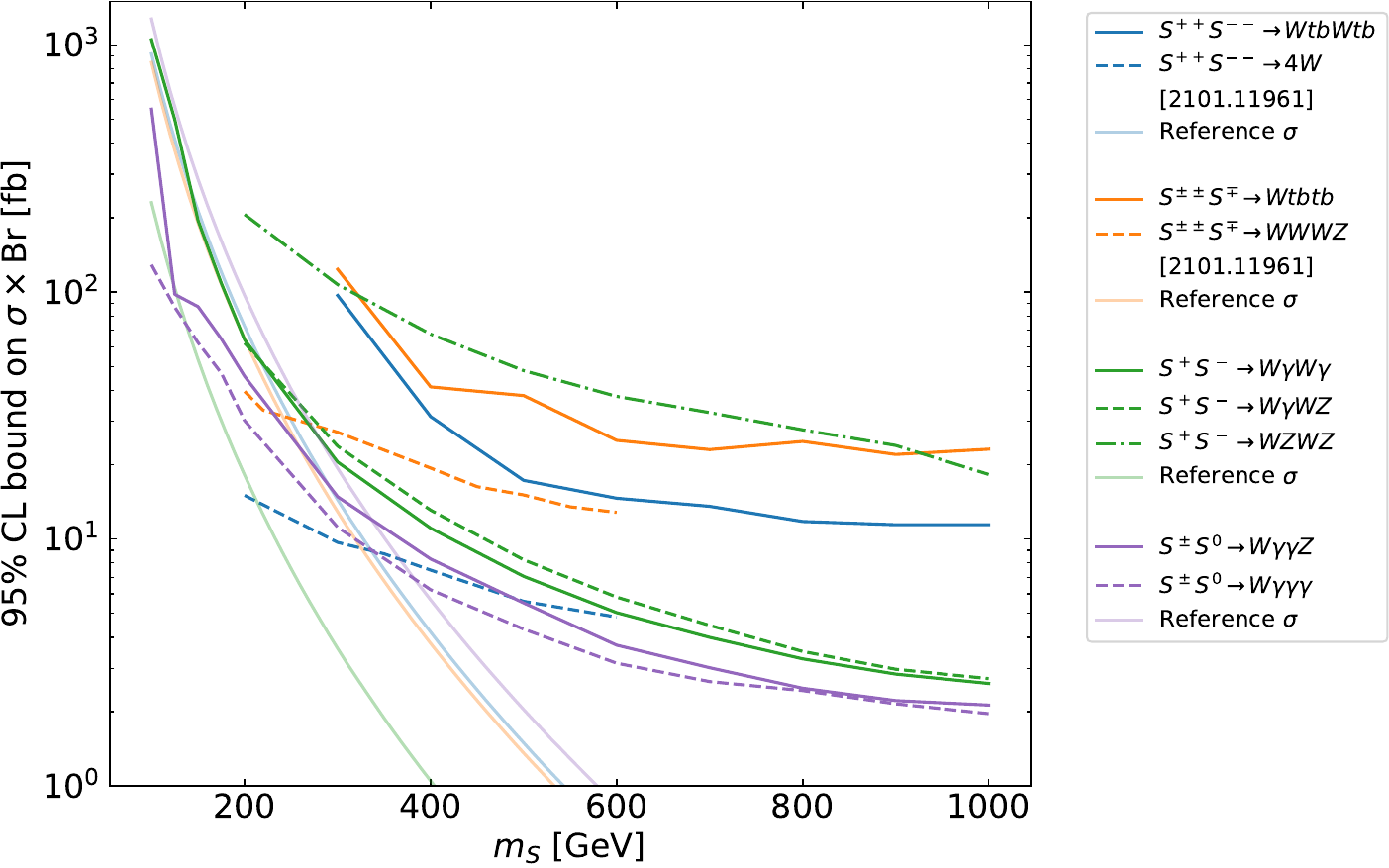}
	\caption{Upper limits on the cross section times branching ratio of the Drell-Yan production of electroweak pNGBs.  The bounds are obtained from recasts implemented in \texttt{Contur} and \texttt{MadAnalysis5} except for the ones presented in \cite{ATLAS:2021jol}. The reference cross sections $\sigma$ for pair production are calculated for a custodial quintuplet from a benchmark model that is discussed in detail in \cref{sec:casestudy}.}
	\label{fig:dyquarkscombined}
\end{figure}

The direct production of EW
pNGBs can occur either via Drell-Yan processes or via vector boson fusion.
In case of pair production the Drell-Yan processes
\begin{equation}
\label{eq:DY}
pp \to  S_i^{\pm\pm} S_j^{\mp}\,,\, S_i^{\pm} S_j^{0} \,,\,  
S_i^{++} S_j^{--} \,,\,S_i^{+} S_j^{-}
 \,,\,S_i^{0} S_j^{0} 
\end{equation}
clearly dominate.
The vector boson fusion processes also allow  single production via the anomaly induced couplings, see \cref{LpiVVtilde} in the appendix. However, the corresponding cross sections are rather small as these couplings are loop induced. Nevertheless, 
these couplings become important for the pNGB decays
if couplings to SM fermions are either forbidden or
strongly suppressed. In this case the dominant decays
are
\begin{subequations}
	\begin{align}
		S^{++}_i &\to W^+ W^+ \\
		S^+_i &\to W^+ \gamma, \, W^+Z \\
		S^0_i &\to W^+ W^-, \, \gamma \gamma, \, \gamma Z, \, ZZ.
	\end{align}
\end{subequations}
Combining these with the various 
production channels leads to a plethora of final states  containing 4 gauge bosons, see for example left diagram in \cref{fig:pNGBpairgraphs}. 
A few of these processes have been searched for directly:
Ref.~\cite{ATLAS:2021jol} sets bounds on the cross section times branching ratio of $S^{++} S^{--} \to WWWW$ and $S^{\pm\pm} S^\mp \to WWW Z$, which are included in \cref{fig:dyquarkscombined} .

Given a coset, the anomaly couplings of pNGBs to vector bosons are fixed up to an overall constant, while the couplings to fermions are more model-dependent.
It is typically possible to construct both fermiophobic as well as fermiophilic models.
If present, the couplings of pNGBs to quarks in \cref{L_ffpi} of the appendix are expected to scale with $m_q/f$, i.e.\
\begin{equation}
	\kappa_t^{S_i^0} \sim \frac{m_t}{f}, \qquad \kappa_b^{S_i^0} \sim \frac{m_b}{f}, \qquad \kappa_{tb,L/R}^{S_i^+} \sim \frac{m_t}{f},
\end{equation}
up to factors of order 1, where $f$ is the pNGB decay constant.
In this case the branching ratios into third-generation quarks dominate over the loop-induced anomaly decays into vector bosons. The corresponding
decay channels are
\begin{subequations}
	\begin{align}
		S^{++} &\to W^+ t\bar b, \\
		S^+ &\to t\bar b, \\
		S^0 &\to t\bar t, \, b\bar b.
	\end{align}
\end{subequations}
Note that the three-body decay $S^{++} \to W^+t\bar b$ proceeds via an off-shell $S^+$, see right diagram in \cref{fig:pNGBpairgraphs}. 
To our knowledge, this channel has not been studied in the literature before.
The corresponding branching ratio is usually close to 100\% as the decay $S^{++} \to W^+ W^+$ is loop-induced. 
In this section we assume that all pNGBs have the same mass. Effects induced due to mass differences
are more model-dependent, hence we defer the 
corresponding discussion to  \cref{sec:casestudy}
for a specific model.

\begin{table}
	\centering
	\ra{1.2}
	\begin{tabular}{lll}\toprule
		\texttt{Contur} pool &  Description &
		final state(s) \\\midrule
		\texttt{ATLAS-13-LL-GAMMA} &  dilepton and $\geq 1$ photon \cite{ATLAS:2019gey}  & $W\gamma W\gamma$, \\
		\texttt{ATLAS-13-GAMMA} &  inclusive multiphotons \cite{ATLAS:2021mbt} & $W\gamma W\gamma$, $W \gamma W Z$ \\
		& & $W\gamma \gamma\gamma$ \\
		\texttt{ATLAS-13-GAMMA-MET} & photon and MET \cite{ATLAS:2018nci} &  $W \gamma W Z$ \\
		\texttt{ATLAS-13-4L} & four leptons ~\cite{ATLAS:2021kog} & $W Z W Z$\\
		\texttt{ATLAS-13-L1L2METJET} & unlike dilepton, MET and jets \cite{ATLAS:2021jgw} & $W\gamma W\gamma$, $W Z W Z$ \\
		\texttt{ATLAS-13-MMJET} & $\mu^+\mu^-$ at the $Z$ pole, plus optional jets~\cite{ATLAS:2019ebv} & $W Z W Z$\\
		\texttt{CMS-13-EEJET} & $e^+e^-$ at the $Z$ pole, plus optional jets~\cite{CMS:2019raw} & $W Z W Z$\\
		\texttt{CMS-13-MMJET} & $\mu^+\mu^-$ at the $Z$ pole, plus optional jets~\cite{CMS:2019raw} & $W Z W Z$ \\
		\bottomrule
	\end{tabular}
	\caption{\texttt{Contur} pools that are most sensitive to the final states listed in the last column.}
	\label{tab:conturpools}
\end{table}

We determine the extent to which the currently recasted analyses can constrain processes for which no direct searches exist.
To this end, we have implemented simplified models  in \texttt{FeynRules} \cite{Alloul:2013bka} at  leading order.
We have generated $10^5$ events using \texttt{MadGraph5\_aMC@NLO} \cite{Alwall:2014hca} version 3.3.1 with the \texttt{NNPDF 2.3} set of parton densities \cite{Ball:2012cx,Buckley:2014ana} and hadronized them with \texttt{Pythia8} \cite{Sjostrand:2014zea}. 
To obtain constraints on the parameter space, the events are analysed with  \texttt{MadAnalysis5},  \cite{Conte:2012fm,Conte:2014zja,Dumont:2014tja,Conte:2018vmg} version 1.9.60.  
\texttt{MadAnalysis5} performs an event reconstruction using \texttt{Delphes 3} \cite{deFavereau:2013fsa} and the anti-$k_T$ algorithm \cite{Cacciari:2008gp} implemented in \texttt{FastJet} \cite{Cacciari:2011ma}, and calculates an exclusion using the CL$_s$ prescription \cite{Read:2002hq}.
We also tested the generated events against the SM measurements implemented in \texttt{Rivet} \cite{Bierlich:2019rhm} version 3.1.5 and have calculated exclusions from the respective \texttt{YODA} files with \texttt{Contur} \cite{Butterworth:2019wnt, Buckley:2021neu}.
Here, we have used all available analyses in \texttt{Contur} version 2.1.1, as well as Refs.~\cite{ATLAS:2021mbt, ATLAS:2018nci, ATLAS:2019cbr} from the upcoming release 2.2.1.

For final states with quarks (in particular top quarks) we found BSM searches implemented in  \texttt{MadAnalysis5} to be the most constraining.\footnote{An analysis using BSM searches implemented in \texttt{CheckMATE} \cite{Drees:2013wra,Dercks:2016npn} is in preparation.} In contrast, in case of vector boson final states, we obtain the  bounds in general from measurements implemented in \texttt{Rivet}, on which
\texttt{Contur} is based.
In \cref{fig:dyquarkscombined} we present the
bounds  on the pair-production cross section times branching ratio for various final states. Beyond our recast results, we also  included the results
of the ATLAS search for $pp\to S^{++} S^{--}\to WWWW$ and
$pp\to S^{\pm\pm}S^{\mp} \to WWWZ$ \cite{ATLAS:2021jol} for completeness. In
\cref{tab:conturpools} we give the most sensitive \texttt{Contur} pools for the various vector boson final states. In the case of quark final states the strongest bounds stem from the four-top search \texttt{CMS-TOP-18-003} \cite{CMS:2019rvj}, which is implemented in \texttt{MadAnalysis5} \cite{Darme:2020hxc}.

For reference, in \cref{fig:dyquarkscombined}
we also show the production cross section for members
of a custodial quintuplet, which is part of the benchmark scenario based on the $SU(5)/SO(5)$ coset discussed in \cref{sec:casestudy}. Note that the references are production cross sections, while the bounds shown are constraints on $\sigma\times\mathrm{Br}$. To obtain bounds on $m_S$, these reference cross sections must be multiplied by the (model-dependent) branching ratios into the respective final states. 
Comparing the theoretical cross sections to the obtained constraints, it is clear that the currently implemented analyses can give bounds on $m_S$ of at most 400 GeV (in the $S^\mp S^0$-channel, if the respective branching ratios to $W\gamma$ and $\gamma\gamma$ are close to 1). We want to stress that  other models give cross section of the same size. For example, the cross sections shown for $S^{++} S^{--}$ and $S^{\pm\pm} S^\mp$ are the same as for the $SU(2)_L$ triplet
used in the ATLAS analysis \cite{ATLAS:2021jol}. The main model dependence enters not in the pair production cross section but in the branching ratios of the pNGB decays.

\subsection{Vector-like quarks}

\begin{figure}[t]
    \centering
\includegraphics{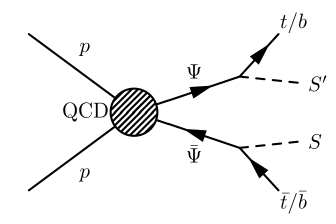} 
    \caption{Illustration of VLQ pair production with subsequent decays into pNGBs and third generation quarks.}
    \label{fig:VLQpairgraphs}
\end{figure}
Coloured particles have a privileged role in the determination of model-independent constraints from searches at the LHC, since the cross-section of QCD pair production depends exclusively on their masses. VLQs have been extensively searched at the LHC by both ATLAS and CMS, under the general assumption that the SM is minimally extended with one VLQ, directly decaying into SM final states. Strong bounds on their masses (above the \TeV) have been obtained regardless of their decay channels~\cite{CMS:2017ynm,ATLAS:2018alq,ATLAS:2018cye,ATLAS:2018mpo,ATLAS:2017nap,ATLAS:2018tnt,ATLAS:2018uky,ATLAS:2017vdo,ATLAS:2018ziw,CMS:2018zkf,CMS:2019kaf,CMS:2020ttz}. 

We have seen, however, that in the scenarios at hand VLQs can also interact with further lighter exotic states, such as new pNGBs, changing their decay patterns as indicated in \cref{fig:VLQpairgraphs}. Exclusion bounds obtained under the aforementioned hypotheses thus require a reinterpretation to account for the reduced branching ratios into SM final states. Moreover, the new decay channels can be tested against experimental data to obtain combined limits on the masses of VLQs and the new states they can decay into. Experimental searches targeting exotic decays of VLQs are not available yet, however phenomenological analyses providing recasts of different searches have already appeared~\cite{Banerjee:2016wls,Chala:2018qdf,Xie:2019gya,Benbrik:2019zdp,Ramos:2019qqa,Aguilar-Saavedra:2019ghg,Brooijmans:2020yij,Wang:2020ips,Dermisek:2020gbr,Dermisek:2021zjd,Roy:2020fqf}. 

\begin{figure}[th]
	\centering
	\includegraphics[width=0.75\textwidth]{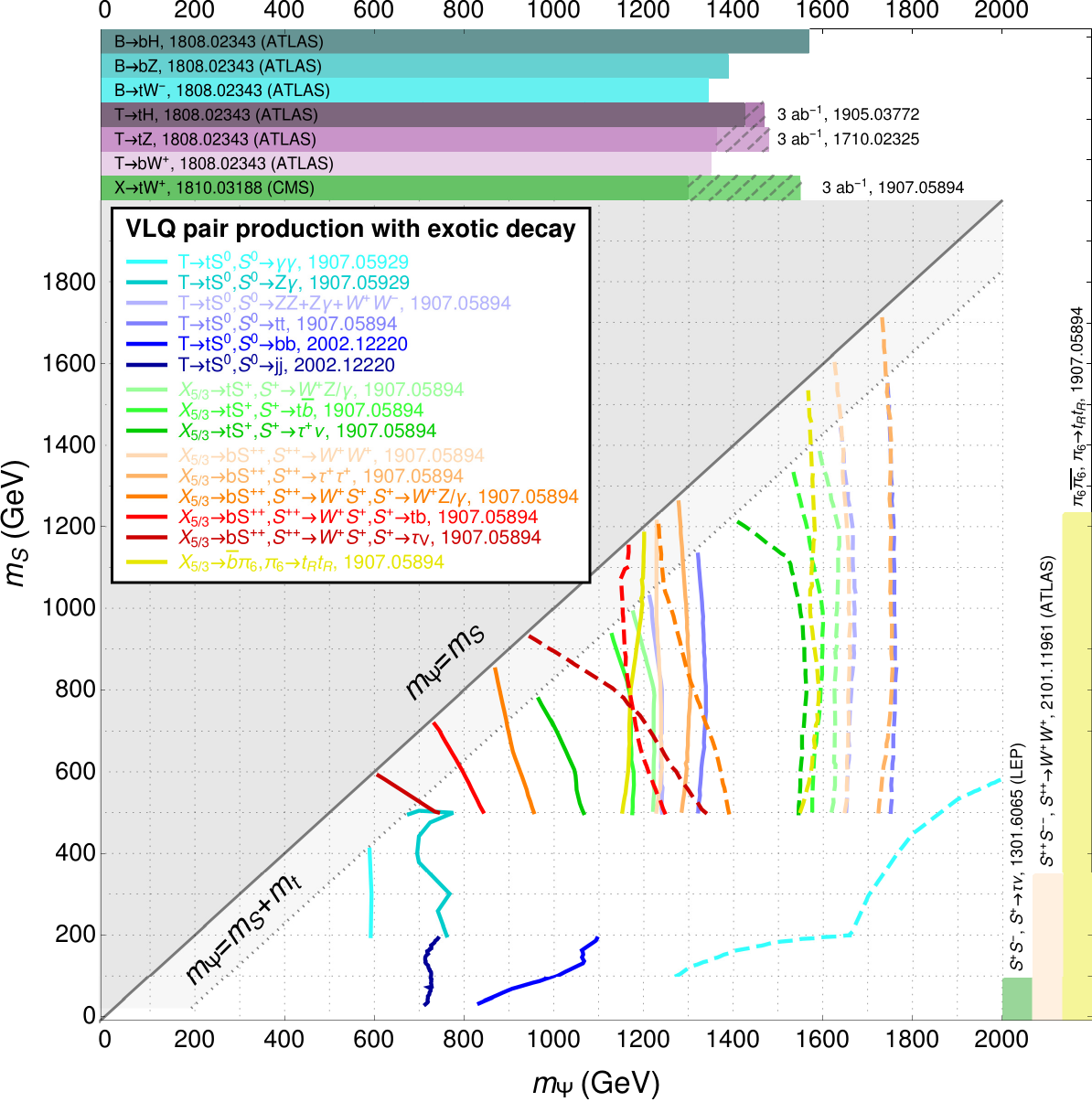}
	\caption{95\% CL limits in the $\{m_\Psi,m_S\}$ plane for VLQs decaying to new pNGBs. Current bounds (13 TeV, 36 fb$^{-1}$) are shown by the solid lines while the projections for the high luminosity LHC (3 ab$^{-1}$) are shown in dashed lines (for details, see corresponding references). Current experimental bounds on the masses of VLQs decaying to SM particles only and the masses of spin-0 states produced in pairs are provided as horizontal and vertical bars respectively. The hatched regions in the bars denote projections for 3 ab$^{-1}$ luminosity.}
	\label{fig:VLQSummaryPlot}
\end{figure}

A summary of results from some of these analyses, corresponding to exclusive decays (100\% Br) of the VLQs into different exotic states, is presented in \cref{fig:VLQSummaryPlot}. When additional decay channels are open, bounds on VLQ masses can be significantly lower than those reported in current experimental searches. In this case, results from  \cref{fig:VLQSummaryPlot} should be accompanied by maps of experimental efficiencies or upper limits on the cross-sections in the same plane for each relevant decay channel of the VLQ.\footnote{An explicit example of this kind of analysis has been proposed in~\cite{Corcella:2021mdl} for pair production of $X_{5/3}$ decaying to light leptons, not motivated by composite scenarios.} The representation proposed in \cref{fig:VLQSummaryPlot} has thus the purpose to provide an overview and a general reference for where realistic bounds can be placed. We also present the projected exclusions in the same channels for high luminosity LHC (3 ab$^{-1}$) \cite{Xie:2019gya,Benbrik:2019zdp}.

To allow an easy comparison with experimental results, current bounds (and  future projections for some cases) on the masses of VLQs decaying to SM particles and bounds on the masses of pair-produced pNGBs have also been included as horizontal \cite{ATLAS:2018ziw,CMS:2018ubm,Barducci:2017xtw,Li:2019wpa} and vertical \cite{ALEPH:2013htx,ATLAS:2017xqs,ATLAS:2021jol} bars respectively. Notice that except for the colour sextet $\pi_6$, a residual model dependence affects the exclusion limits: the pair production cross-section of an EW pNGB depends on its coupling to the $Z$ boson, which is associated with its representation under the EW gauge group. Experimental bounds on charged scalars are obtained under specific assumptions, for which we refer to the corresponding literature~\cite{ALEPH:2013htx,ATLAS:2017xqs,ATLAS:2021jol}.

\section{Case study: $SU(5)/SO(5)$} 
\label{sec:casestudy}

As a concrete example, we now consider a model based on the $SU(5)/SO(5)$ coset after having integrated out the VLQs. The UV theory can be constructed as in~\cite{Ferretti:2014qta} and it's being studied on the lattice~\cite{Ayyar:2018zuk,Ayyar:2018glg}.
This coset delivers a total of 14 pNGBs, which include the Higgs bidoublet, a singlet $\eta$, and a bitriplet under $SU(2)_L\times SU(2)_R$. Here we focus on the bitriplet, as direct $\eta$ production at the LHC is strongly suppressed. The bitriplet
decomposes as a quintuplet $\eta_5\equiv (\eta^0_5, \eta^\pm_5, \eta^{\pm\pm}_5)$, a triplet $\eta_3\equiv (\eta^0_3, \eta^\pm_3)$ and a singlet $\eta_1^0$ of custodial $SU(2)_C$.

\begin{figure}[t]
\begin{center}
    	\includegraphics[width=0.35\linewidth]{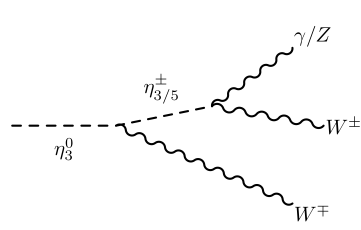} \hspace{30pt}
    		\includegraphics[width=0.35\linewidth]{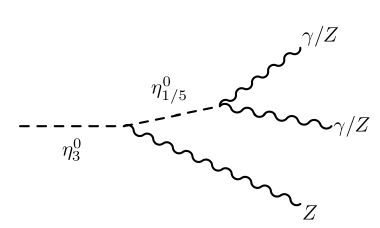} \vspace{2ex}
\end{center}    	
    \caption{Feynman diagrams corresponding to three-body decays of $\eta_3^0$.}
    \label{fig:decays_eta30}
\end{figure}

A detailed description of how all the couplings can be derived, including the VLQ sector, has been given in~\cite{Banerjee:2022izw}. For this section we consider the exotic pNGBs to be fermiophobic. In Appendix~\ref{app:SU5SO5}, we collect the relevant couplings, which depend on the pNGB decay constant $f$, which we set to 1 TeV. The remaining parameters to be chosen are the masses for the custodial multiplets $m_{1,3,5}$~\footnote{In principle members of different custodial multiplets can mix as discussed in \cite{Agugliaro:2018vsu}.}. Note that the exotic pNGBs do not acquire a vacuum expectation value in this model.

\begin{figure}[t]
	\centering
	\includegraphics[width=0.45\linewidth]{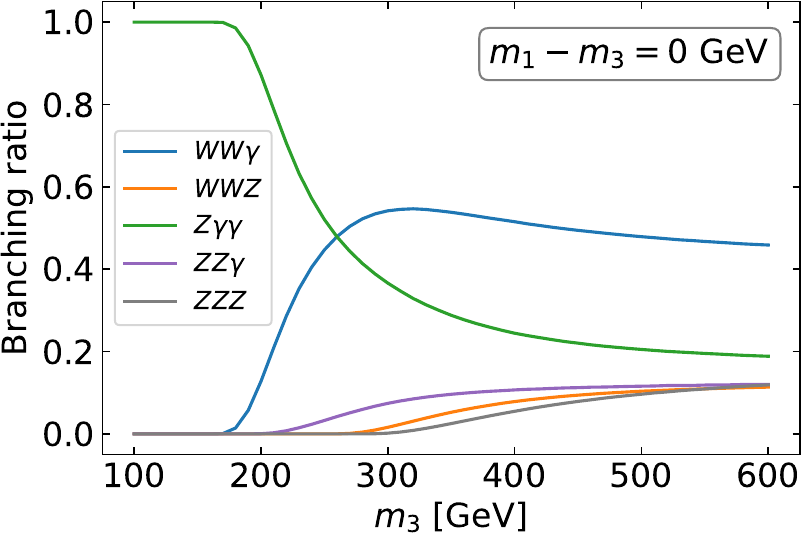} \quad
	\includegraphics[width=0.45\linewidth]{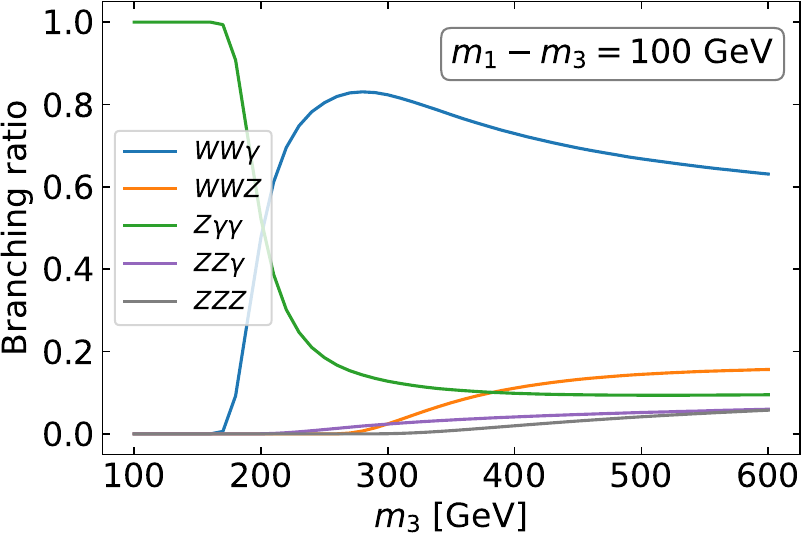} 
	\caption{Three-body decays of $\eta_3^0$ for the scenario $m_{1,3}\ll m_5$. When kinematically accessible, the dominant channel is $\eta_3^0\to W^\pm \eta_3^{\mp (*)}$. 
	}
	\label{fig:xsetaa}
\end{figure}

\begin{figure}[]
	\centering
	\includegraphics[width=.45\linewidth]{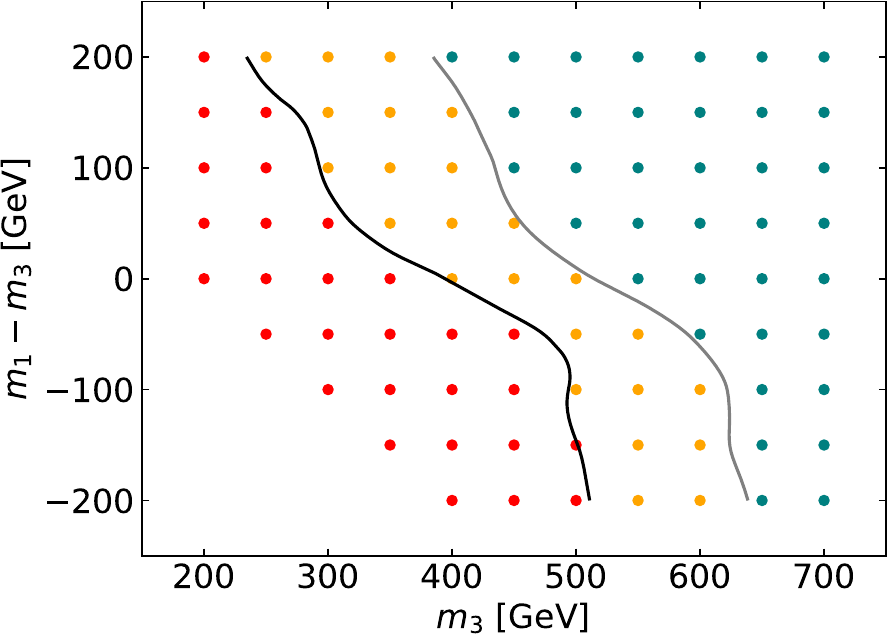}\quad
	\includegraphics[width=.45\linewidth]{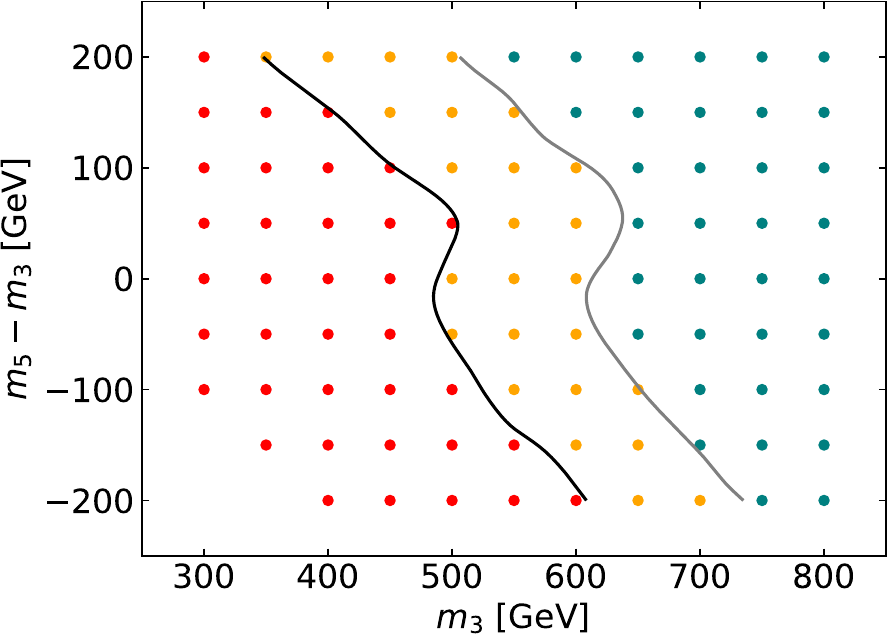}
	\caption{Exclusion bounds on the pNGB masses in $m_3 - \Delta m$ plane, where $\Delta m = m_1-m_3$ (left panel) and $\Delta m = m_5 - m_3$ plane (right panel). In both cases we assume the pNGBs are fermiophobic and produced in pair through Drell-Yan processes. The black (grey) lines represent the 2-$\sigma$ (1-$\sigma$) exclusion. Red points are excluded with CL above 2-$\sigma$ and yellow ones between 1 and 2-$\sigma$.
	}
	\label{fig:splittingplotb}
\end{figure}

In this scenario, these pNGBs are mainly produced through  Drell-Yan processes (see \cref{eq:DY}).
In the presence of a mass splitting between the custodial multiplets the decay pattern changes with respect to the discussion in \cref{sec:colliderpheno}. In such case
the lowest lying multiplet will still decay only
via the anomaly. However, the heavier pNGBs will
decay dominantly into a possibly off-shell vector boson and a member
of the lower lying multiplet, as soon as the mass splitting is larger than about 1 GeV. As an example we take
$m_3 \ge m_{5,1}$, in which case the $\eta_3$ members decay dominantly as
\begin{align}
\eta_3^+ &\to \eta_5^{++}  W^{-(*)} \,,\,\, \eta_5^{+} Z^{(*)} \,,\, \eta^0_5 W^{+(*)} \,,\, \eta^0_1 W^{+(*)}\,;\\
\eta_3^0 & \to \eta_5^{\pm} W^{\mp(*)} \,,\,\, \eta_5^0 Z^{(*)}\,,\,\, \eta_1^0 Z^{(*)}\,. \nonumber
\end{align}
Since $\Gamma_{\eta_{5,1}} \ll \Gamma_{W,Z}$, the decays into off-shell $\eta_{5,1}$ and on-shell vector bosons are strongly suppressed.
Note that in scenarios where $\eta_3$ is the lightest multiplet, $\eta^0_3$ is forced to have three body decays since it does not couple to the anomaly, see \cref{fig:decays_eta30}. We give examples of branching ratios of $\eta^0_3$ for different scenarios in \cref{fig:xsetaa}.
We use the same set-up as in \cref{sec:colliderpheno} to determine bounds at the LHC, assuming a global $K$-factor of $1.15$ to take QCD corrections into account. 
We find again that \texttt{Contur} yields stronger bounds than currently available analyses in \texttt{MadAnalysis5}. The leading bounds are displayed in \cref{fig:splittingplotb} for two limiting scenarios: $m_{1,3} \ll m_5$ (left plot) and $m_{3,5} \ll m_1$ (right plot). The bounds on the right side are much stronger due to the presence of the additional singly and doubly charged states in the quintuplet.

\section*{Acknowledgements}

We thank Jon Butterworth for helpful discussions on \texttt{Contur}. 
A.B.,\ D.B.F.,\ G.F.\ and L.P.\ acknowledge support from the Knut and Alice Wallenberg foundation under the grant KAW
2017.0100 (SHIFT project). 
G.C., A.D., M.K, W.P.\ and L.S. have been supported by the ``DAAD, Frankreich'' and ``Partenariat Hubert Curien (PHC)'' PROCOPE 2021-2023, project number 57561441.
T.F,\ M.K.,\ W.P\ and L.S were supported under the framework of the international cooperation program ``GenKo'' managed by the National Research Foundation of Korea (No. 2022K2A9A2A15000153, FY2022) and DAAD, P33 - projekt-id 57608518.
G.C., A.D., T.F. and B.F. acknowledge support from the Campus-France STAR project n. 43566UG, ``Higgs and Dark Matter connections''.
T.F.\ is supported by
KIAS Individual Grant (AP083701) via the Center for AI and Natural Sciences at Korea Institute for Advanced Study.
G.C. and A.D. are grateful to the LABEX Lyon Institute of Origins (ANR-10-LABX-0066) Lyon for its financial support within the program ``Investissements d'Avenir'' of the French government operated by the National Research Agency (ANR).

\appendix

\section{The general Lagrangian for models with VLQs and pNGBs}
\label{lagrangian}

In this appendix we provide the explicit expressions of each of the terms shown in \cref{L_int} which is also implemented in the software tools, albeit with a limited particle content. The terms $\mathcal{L}_{SS V}$, $\mathcal{L}_{SS VV}$, and $\mathcal{L}_{\pi\pi V+\pi\pi VV}$ are simply the couplings of the pNGBs with the gauge bosons, arising from the covariant derivative. 
\begin{align}
\mathcal{L}_{SS V} &=\frac{ie}{s_W}W^{-\mu}\sum_{i,j}\left[K^{S^0_i S^+_j}_{W}S^0_i\overleftrightarrow{\partial_\mu}S^+_j+K^{S^-_i S^{++}_j}_{W}S^-_i\overleftrightarrow{\partial_\mu}S^{++}_j\right]+{\mathrm{h.c.}} \nonumber\\
& +\frac{ie}{s_Wc_W}Z^\mu \left[\sum_{i<j} K^{S^0_i S^0_j}_{Z} S^0_i\overleftrightarrow{\partial_\mu} S^0_j+ \sum_{i,j} \left(K^{ S^+_i S^{-}_j}_{Z} S^+_i\overleftrightarrow{\partial_\mu} S^{-}_j+K^{ S^{++}_i S^{--}_j}_{Z} S^{++}_i\overleftrightarrow{\partial_\mu} S^{--}_j \right)\right]\nonumber\\
&-ieA^\mu \sum_{i}\left[ S^+_i\overleftrightarrow{\partial_\mu} S^-_i+2 S^{++}_i\overleftrightarrow{\partial_\mu} S^{--}_i\right],
\label{L_pipiV} 
\end{align}
\begin{align}
\mathcal{L}_{ S S VV} &=e^2 A_\mu A^\mu \sum_{i}\left[  S^+_i S^-_i+4 S^{++}_i S^{--}_i\right] \nonumber\\ 
&+\frac{e^2}{s_Wc_W}A^\mu Z_\mu  \sum_{i,j}\left[ K^{ S^+_i S^-_j}_{AZ} S^+_i S^-_j+ K^{ S^{++}_i S^{--}_j}_{AZ} S^{++}_i S^{--}_j \right] \nonumber\\
&+\frac{e^2}{s^2_Wc^2_W} Z_\mu Z^\mu \left[\sum_{i\leq j}K^{ S^0_i S^0_j}_{ZZ} S^0_i S^0_j+ \sum_{i,j} \left( K^{ S^+_i S^-_j}_{ZZ} S^+_i S^-_j+ K^{ S^{++}_i S^{--}_j}_{ZZ} S^{++}_i S^{--}_j \right) \right] \nonumber\\
&+\frac{e^2}{s^2_W}W^+_\mu W^{-\mu} \left[\sum_{i\leq j} K^{ S^0_i S^0_j}_{WW} S^0_i S^0_j+ \sum_{i,j} \left( K^{ S^+_i S^-_j}_{WW} S^+_i S^-_j+ K^{ S^{++}_i S^{--}_j}_{WW} S^{++}_i S^{--}_j \right) \right] \nonumber\\
&+\frac{e^2}{s_W} A^\mu W^-_\mu  \sum_{i,j} \left[ K^{ S^0_i S^+_j}_{AW^-} S^0_i S^+_j+ K^{ S^-_i S^{++}_j}_{AW^-} S^-_i S^{++}_j\right]+{\mathrm{h.c.}} \nonumber\\
&+\frac{e^2}{s^2_Wc_W} Z^\mu W^-_\mu \sum_{i,j}\left[ K^{ S^0_i S^+_j}_{ZW^-} S^0_i S^+_j+ K^{ S^-_i S^{++}_j}_{ZW^-} S^-_i S^{++}_j\right]+{\mathrm{h.c.}} \nonumber\\
&+\frac{e^2}{s^2_W} W^-_\mu W^{-\mu} \left[\sum_{i\leq j} K^{ S^+_i S^+_j}_{W^-W^-} S^+_i S^+_j+ \sum_{i,j} \left(K^{ S^{0}_i S^{++}_j}_{W^-W^-} S^{0}_i S^{++}_j \right)\right]+{\mathrm{h.c.}},
\label{L_pipiVV}
\end{align}
\begin{align}
\mathcal{L}_{\pi\pi V+\pi\pi VV} & = 
\sum_{r=3,6,8} (D_\mu \pi^{q}_r )^\dagger D^\mu \pi^{q}_r,
\end{align}
with $ D_\mu =\partial_\mu - i g_s  G^{a\mu} T^a_r - i e q(A_\mu-\tan\theta_W Z_\mu)$ and $s_W (c_W)\equiv \sin\theta_W (\cos\theta_W)$. Note that  the colored pNGBs do not carry any $SU(2)_L$ index in the models considered here and, thus, the couplings to the $Z$ arise purely from hypercharge.
The $\mathcal{L}_{ S VV}$ consists of the coupling of the Higgs boson to the EW gauge bosons.
\begin{align}
\mathcal{L}_{ S VV} =2M_W^2 \frac{h}{v}\left[K^{h}_{WW} W_\mu^+W^{-\mu}+K^{h}_{ZZ} \frac{1}{2c_W^2}Z_\mu Z^\mu\right]\,.  \label{L_piVV}
\end{align}	
Any additional pNGB acquiring a vacuum expectation value contributes to the masses of the gauge bosons and acquires a coupling of this type. We consider here the case where only the Higgs doublet receives a vacuum expectation value ($v$). In this case, the neutral scalars easily evade all LEP collider bounds and no corrections to the tree-level $\rho$-parameter arise. This choice is well motivated by a study of the potential for models given in~\cite{Ferretti:2016upr,Agugliaro:2018vsu}. 

The Lagrangian $\mathcal{L}_{ S V\tilde{V}}$ and $\mathcal{L}_{\pi V\tilde{V}}$  contain the dimension five anomalous couplings of the pNGBs with the vector boson field strengths. If one proceeds to integrate out the third family or the heavy vector bosons, both $\mathcal{L}_{ S VV}$ and $\mathcal{L}_{ S V\tilde{V}}$ include additional dimension five couplings such as the coupling of the Higgs boson to the gluons.
\begin{align}
\nonumber
\mathcal{L}_{ S V\tilde{V}} &= \frac{e^2}{16 \pi^2 v} \bigg[ \sum_i S^0_i\left(\tilde{K}^{ S^0_i}_{\gamma\gamma}F_{\mu\nu}\tilde{F}^{\mu\nu}+\frac{2}{s_Wc_W}\tilde{K}^{ S^0_i}_{\gamma Z}F_{\mu\nu}\tilde{Z}^{\mu\nu}+\frac{1}{s_W^2c_W^2}\tilde{K}^{ S^0_i}_{ZZ}Z_{\mu\nu}\tilde{Z}^{\mu\nu}\right.\nonumber\\
&\left.+\frac{2}{s_W^2}\tilde{K}^{ S^0_i}_{WW}W^+_{\mu\nu}\tilde{W}^{-\mu\nu}\right)
+\sum_i S^+_i \left( \frac{2}{s_W}\tilde{K}^{ S^+_i}_{\gamma W}F_{\mu\nu}\tilde{W}^{-\mu\nu}+ \frac{2}{s_W^2c_W}\tilde{K}^{ S^+_i}_{ZW}Z_{\mu\nu}\tilde{W}^{-\mu\nu}\right)+{\rm h.c.} \nonumber\\
&+ \frac{1}{s_W^2}\sum_i S^{++}_i \tilde{K}^{ S^{++}_i}_{W^-W^-}W^-_{\mu\nu}\tilde{W}^{-\mu\nu}+{\rm h.c.}\bigg],
\label{LpiVVtilde}
\end{align}
\begin{align}
\mathcal{L}_{\pi V\tilde V} & =  \frac{1}{16 \pi^2 f} \left(g_s^2\tilde K^{\pi^0_8}_{gg} \pi^0_8 G_{\mu\nu} \tilde G^{\mu\nu} +
2g_s e\tilde{K}^{\pi^0_8}_{g\gamma} \pi^0_8 G_{\mu\nu} \tilde F^{\mu\nu} +
\frac{2g_s e}{s_W c_W}\tilde{K}^{\pi^0_8}_{gZ} \pi^0_8 G_{\mu\nu} \tilde Z^{\mu\nu} \right).
\end{align}
The only terms in $\mathcal{L}_{\Psi \Psi V}$ relevant for VLQ pair production are their couplings to the gluons
\begin{align}
\mathcal{L}_{\Psi \Psi V} &=\frac{g_s}{2}\left[\bar{T}\slashed G^a\lambda^a T+\bar{B}\slashed G^a\lambda^a B+\bar{X}\slashed G^a\lambda^a X\right],
\label{L_psipsiV}
\end{align}
with $\lambda^a$ being the usual Gell-Mann matrcies.
The terms $\mathcal{L}_{\Psi f V}$, $\mathcal{L}_{\Psi f  S}$, and $\mathcal{L}_{\Psi f  \pi}$ arise from partial compositeness by linearly coupling the VLQs ($\Psi$) to the third quark family ($f$). 
\begin{align}
\mathcal{L}_{\Psi f V} &=\frac{e}{\sqrt{2}s_W}\kappa^W_{T,L}\bar{T}\slashed W^+ P_L b+\frac{e}{2c_Ws_W}\kappa^Z_{T,L}\bar{T}\slashed Z P_L t+\frac{e}{\sqrt{2}s_W}\kappa^W_{B,L}\bar{B}\slashed W^- P_L t \nonumber\\
&+\frac{e}{2c_Ws_W}\kappa^Z_{B,L}\bar{B}\slashed Z P_L b+\frac{e}{\sqrt{2}s_W}\kappa^W_{X,L}\bar{X}\slashed W^+ P_L t+ L\leftrightarrow R +{\mathrm{h.c.}}
\label{L_psifV}
\end{align}
\begin{align}
\mathcal{L}_{\Psi f  S} & =\sum_{i}  S^+_i\left[\kappa^{ S^+_i}_{T,L} \bar{T} P_L b + \kappa^{ S^+_i}_{X,L} \bar{X} P_L t+ L\leftrightarrow R\right]+ {\rm h.c.} + \sum_{i}  S^-_i\left[\kappa^{ S^-_i}_{B,L} \bar{B} P_L t + L\leftrightarrow R\right]+ {\rm h.c.} \nonumber\\
&+\sum_{i} S^0_i\left[\kappa^{ S^0_i}_{T,L}\bar{T} P_L t+\kappa^{ S^0_i}_{B,L}\bar{B} P_L b+ L\leftrightarrow R\right]+ {\rm h.c.}\nonumber \\
&+\sum_{i}  S^{++}_i\left[\kappa^{ S^{++}_i}_{X,L} \bar{X} P_L b + L\leftrightarrow R\right]+ {\rm h.c.}
\label{L_psifpi}
\end{align}
\begin{align}
\nonumber
\mathcal{L}_{\Psi f \pi} & = \pi^0_8 
\left[ \kappa^{\pi^0_8}_{T,L} \bar{T} P_L t
+ \kappa^{\pi^0_8}_{B,L} \bar{B} P_L b 
 + L\leftrightarrow R \right]+ {\rm h.c.}
\nonumber \\
& +\left[ \pi^{4/3}_6 \left(\kappa^{\pi^{4/3}_6}_{T,L}  \bar{T} C P_R \bar{t}^T  + \kappa^{\pi^{4/3}_6}_{X,L} \bar{X} C P_R \bar{b}^T \right)  + \pi^{2/3}_6 \kappa^{\pi^{2/3}_6}_{B,L} B^T C P_L b   + L\leftrightarrow R \right]+ {\rm h.c.} 
\end{align}
with $C$ being the charge conjugation operator. The terms $\mathcal{L}_{ff S}$ and $\mathcal{L}_{f f \pi}$ involving two SM third generation quarks are given below with $y_f=\sqrt{2}m_f/v$. 
\begin{align}
\mathcal{L}_{f f  S} & =-\frac{h}{\sqrt{2}}\left[y_t \kappa^h_{t} \bar{t}t+y_b \kappa^h_{b} \bar{b}b\right]+\sum_{i} S^0_i \bigg[\bar{t}\left(\kappa^{ S^0_i}_{t}+i\tilde{\kappa}^{ S^0_i}_{t}\gamma_5\right)t+\bar{b}\left(\kappa^{ S^0_i}_{b}+i\tilde{\kappa}^{ S^0_i}_{b}\gamma_5\right)b\bigg] \nonumber \\
& +\sum_{i}  S^+_i\left[\kappa^{ S^+_i}_{tb,L} \bar{t} P_L b+ L\leftrightarrow R\right]+ {\rm h.c.}
\label{L_ffpi}
\end{align}
\begin{align}
\mathcal{L}_{f f \pi} & = \pi^0_8 
\left[  \bar{t} 
(\kappa^{\pi^0_8}_t + i \tilde \kappa^{\pi^0_8}_t \gamma_5) t
+ \bar{b} (\kappa^{\pi^0_8}_b + i \tilde \kappa^{\pi^0_8}_b \gamma_5) b 
  \right] \nonumber \\
 & + \pi^{4/3}_6  \bar{t} (\kappa^{\pi^{4/3}_6}_{t} + i \tilde \kappa^{\pi^{4/3}_6}_{t} \gamma_5) C \bar{t}^T  
 + \pi^{2/3}_6  b^T (\kappa^{\pi^{2/3}_6}_{t} + i \tilde \kappa^{\pi^{2/3}_6}_{t} \gamma_5) C b  + {\rm h.c.} 
\end{align}

\subsection{Couplings in $SU(5)/SO(5)$ coset}
\label{app:SU5SO5}

Now we present the explicit coupling strengths between the weak gauge bosons and EW pNGBs in the custodial basis for $SU(5)/SO(5)$ coset, in terms of $\sin\theta\equiv s_\theta =v/f$. The couplings presented below can be directly used to construct the restriction cards for the Feynrules-based tools we discussed earlier. In \cref{tab:mapping} we map the generic symbols for the scalar fields used in the \cref{L_int} to the specific symbols used for the $SU(5)/ SO(5)$ coset. The coefficient $K^h_{VV}$ ($V=W,Z$), the universal modification of the Higgs coupling with the weak gauge bosons appearing in $\mathcal{L}_{SVV}$ is given by $c_\theta$.

\begin{table}[t!]
	\def\arraystretch{1.3}
	\centering
	\begin{tabular}{ccccccccc}
		\toprule  
		Fields in Eq.~\eqref{L_int} & Fields in section~\ref{sec:casestudy} \\  
		\midrule 
		$S^0_i$ & $h,\eta,\eta^0_1,\eta^0_3,\eta^0_5$ \\
		$S^\pm_i$ & $\eta^\pm_3,\eta^\pm_5$ \\
		$S^{\pm\pm}_i$ & $\eta^{\pm\pm}_5$ \\
		\bottomrule  
	\end{tabular} 
	\caption{ Mapping between the generic symbols for the spin-0 fields to the specific EW pNGBs in the custodial basis for $SU(5)/SO(5)$ coset}
	\label{tab:mapping}
\end{table}

\subsubsection*{Coefficients appearing in $\mathcal{L}_{S S V}$}

\begin{small}
\begin{equation*}
\def\arraystretch{1.4}
\begin{array}{cccccccccc}
\toprule 
&\multicolumn{2}{c}{K^{S^0_iS^+_j}_W} & K^{S^-_iS^{++}_j}_W\\
\midrule  
& \eta^+_3 & \eta^+_5 & \eta_5^{++}\\ 
\midrule 
h &  0 & 0 & \multirow{5}{*}{$-$}\\
\eta^0_3 & -\frac{i}{2} & \frac{c_\theta}{2} &\\ 
\eta^0_5 & -\frac{c_\theta}{2\sqrt{3}} & \frac{i\sqrt{3}}{2} &\\
\eta^0_1 & \sqrt{\frac{2}{3}}c_\theta & 0 &\\
\eta &  0 & 0 &\\
\eta^-_3 & \multicolumn{2}{c}{\multirow{2}{*}{$-$}} & \frac{c_\theta}{\sqrt{2}} \\ 
\eta^-_5 & & & -\frac{i}{\sqrt{2}}\\
\vphantom{--}\\
\bottomrule
\end{array}
\quad
\begin{array}{ccccccccccccc}
\toprule 
&\multicolumn{5}{c}{K^{S^0_iS^0_j}_Z} & \multicolumn{2}{c}{K^{S^+_iS^{-}_j}_Z} & K^{S^{++}_iS^{--}_j}_Z \\
\midrule 
& h & \eta^0_3 & \eta^0_5 & \eta^0_1 & \eta & \eta^-_3 & \eta^-_5 & \eta_5^{--}\\ 
\midrule  
h &  0 & 0 & 0 & 0 & 0 & \multicolumn{2}{c}{\multirow{5}{*}{$-$}} & \multirow{5}{*}{$-$}\\
\eta^0_3 & & 0 & \frac{i c_\theta}{\sqrt{3}} & i\sqrt{\frac{2}{3}}c_\theta & 0\\ 
\eta^0_5 & & & 0 & 0 & 0 \\
\eta^0_1 & & & & 0 & 0 \\
\eta &  & & & & 0 \\
\eta^+_3 & \multicolumn{5}{c}{\multirow{2}{*}{$-$}} & -\frac{c_{2W}}{2} & -\frac{ic_\theta}{2} \\ 
\eta^+_5 & & & & & & & -\frac{c_{2W}}{2} \\
\eta^{++}_5 & \multicolumn{7}{c}{-} & -c_{2W} \\
\bottomrule 
\end{array}
\end{equation*}
\end{small}

\subsubsection*{Coefficients appearing in $\mathcal{L}_{SSVV}$}

\begin{small}
\begin{equation*}
\begin{array}{ccccccccccccc}
\toprule 
&\multicolumn{5}{c}{K^{S^0_iS^0_j}_{ZZ}} & \multicolumn{2}{c}{K^{S^+_iS^{-}_j}_{ZZ}} & K^{S^{++}_iS^{--}_j}_{ZZ} \\
\midrule 
& h & \eta^0_3 & \eta^0_5 & \eta^0_1 & \eta & \eta^-_3 & \eta^-_5 & \eta_5^{--}\\ 
\midrule 
h &  \frac{c_{2\theta}}{8} & 0 & 0 & 0 & 0 & \multicolumn{2}{c}{\multirow{5}{*}{$-$}} & \multirow{5}{*}{$-$} \\
\eta^0_3 & & \frac{3+5c_{2\theta}}{16} & 0 & 0 & 0\\ 
\eta^0_5 & & & \frac{c_{2\theta}}{6} & \frac{3+5c_{2\theta}}{12\sqrt{2}} & -\sqrt{\frac{5}{6}}\frac{s^2_\theta}{2} \\
\eta^0_1 & & & & \frac{15+17c_{2\theta}}{96} & -\sqrt{\frac{5}{3}}\frac{s^2_\theta}{8} \\
\eta &  & & & & -\frac{5s^2_\theta}{16} \\
\eta^+_3 & \multicolumn{5}{c}{\multirow{2}{*}{$-$}} & \frac{3-2s^2_{2W}+c_{2\theta}}{8} & \frac{i c_{2W} c_\theta}{2} \\ 
\eta^+_5 & & & & & & & \frac{c^2_{2W}+c_{2\theta}}{4} \\
\eta^{++}_5 & \multicolumn{7}{c}{-} & c^2_{2W} \\
\bottomrule 
\end{array}
\end{equation*}

\begin{equation*}
\begin{array}{ccccccccccccc}
\toprule  
&\multicolumn{5}{c}{K^{S^0_iS^0_j}_{WW}} & \multicolumn{2}{c}{K^{S^+_iS^{-}_j}_{WW}} & K^{S^{++}_iS^{--}_j}_{WW} \\
\midrule 
& h & \eta^0_3 & \eta^0_5 & \eta^0_1 & \eta & \eta^-_3 & \eta^-_5 & \eta_5^{--}\\ 
\midrule 
h &  \frac{c_{2\theta}}{4} & 0 & 0 & 0 & 0 & \multicolumn{2}{c}{\multirow{5}{*}{$-$}} & \multirow{5}{*}{$-$} \\
\eta^0_3 & & \frac{3+c_{2\theta}}{8} & 0 & 0 & 0\\ 
\eta^0_5 & & & \frac{9+c_{2\theta}}{12} & -\frac{3+5c_{2\theta}}{12\sqrt{2}} & \sqrt{\frac{5}{6}}\frac{s^2_\theta}{2} \\
\eta^0_1 & & & & \frac{15+17c_{2\theta}}{48} & -\sqrt{\frac{5}{3}}\frac{s^2_\theta}{4} \\
\eta &  & & & & -\frac{5s^2_\theta}{8} \\
\eta^+_3 & \multicolumn{5}{c}{\multirow{2}{*}{$-$}} & \frac{3c^2_\theta}{2} & -\frac{i c_\theta}{2} \\ 
\eta^+_5 & & & & & & & \frac{5+c_{2\theta}}{4} \\
\eta^{++}_5 & \multicolumn{7}{c}{-} & c_\theta^2 \\
\bottomrule  
\end{array}
\end{equation*}

\begin{equation*}
\def\arraystretch{1.4}
\begin{array}{cccccccccc}
\toprule  
&\multicolumn{2}{c}{K^{S^+_iS^-_j}_{AZ}} & K^{S^{++}_iS^{--}_j}_{AZ} \\
\midrule 
& \eta^-_3 & \eta^-_5 & \eta_5^{--} \\ 
\midrule 
\eta^+_3 & c_{2W} & i c_\theta & \multirow{2}{*}{$-$} \\ 
\eta^+_5 &  & c_{2W} \\
\eta^{++}_5 & \multicolumn{2}{c}{-} & 4c_{2W} \\
\bottomrule  
\vphantom{--}\\
\vphantom{--}\\
\vphantom{--}\\
\vphantom{--}\\
\end{array}
\quad
\begin{array}{cccccccccc}
\toprule  
&\multicolumn{2}{c}{K^{S^0_iS^+_j}_{AW}} & K^{S^-_iS^{++}_j}_{AW}\\
\midrule 
& \eta^+_3 & \eta^+_5 & \eta_5^{++}\\ 
\midrule 
h &  0 & 0 & \multirow{5}{*}{$-$}\\
\eta^0_3 & -\frac{i}{2} & \frac{c_\theta}{2} \\ 
\eta^0_5 & -\frac{c_\theta}{2\sqrt{3}} & \frac{i\sqrt{3}}{2} \\
\eta^0_1 & \sqrt{\frac{2}{3}}c_\theta & 0 \\
\eta &  0 & 0\\
\eta^-_3 & \multicolumn{2}{c}{\multirow{2}{*}{$-$}} & \frac{3 c_\theta}{\sqrt{2}} \\ 
\eta^-_5  & & & -\frac{3i}{\sqrt{2}}\\
\bottomrule  
\end{array}
\end{equation*}

\begin{equation*}
\def\arraystretch{1.4}
\begin{array}{cccccccccc}
\toprule  
&\multicolumn{2}{c}{K^{S^0_iS^+_j}_{ZW}} & K^{S^-_iS^{++}_j}_{ZW}\\
\midrule  
& \eta^+_3 & \eta^+_5 & \eta_5^{++}\\ 
\midrule  
h &  0 & 0 & \multirow{5}{*}{$-$}\\
\eta^0_3 & \frac{i(s_W^2+c_{2\theta})}{2} &  -\frac{(1+s_W^2)c_{\theta}}{2} \\ 
\eta^0_5 & -\frac{(3-s^2_W)c_\theta}{2\sqrt{3}} & \frac{i(3c_{2W}-c_{2\theta})}{4\sqrt{3}} \\
\eta^0_1 & -\sqrt{\frac{2}{3}}s^2_Wc_\theta & -\frac{i(3+5c_{2\theta})}{4\sqrt{6}} \\
\eta &  0 & \sqrt{\frac{5}{2}}\frac{i s^2_\theta}{2} \\
\eta^-_3 & \multicolumn{2}{c}{\multirow{2}{*}{$-$}} & \frac{(1-3s_W^2 )c_\theta}{\sqrt{2}} \\ 
\eta^-_5  & & & -\frac{i(3c_{2W} -c_{2\theta})}{2\sqrt{2}} \\
\bottomrule  
\end{array}
\quad
\def\arraystretch{1.45}
\begin{array}{cccccccccc}
\toprule  
&\multicolumn{2}{c}{K^{S^+_iS^+_j}_{W^-W^-}} & K^{S^0_iS^{++}_j}_{W^-W^-}\\
\midrule 
& \eta^+_3 & \eta^+_5 & \eta_5^{++}\\ 
\midrule 
\eta^+_3 & -\frac{c_{2\theta}}{4} & \frac{i c_\theta}{2} & \multirow{2}{*}{$-$}\\ 
\eta^+_5 & & \frac{3-c_{2\theta}}{8} \\
h & \multicolumn{2}{c}{\multirow{5}{*}{$-$}} & 0 \\
\eta^0_3 & & & -\frac{ic_\theta}{2} \\ 
\eta^0_5 & & & \frac{3-c_{2\theta}}{2\sqrt{6}}  \\
\eta^0_1 & & & \frac{3+5c_{2\theta}}{8\sqrt{3}} \\
\eta & & &  -\frac{\sqrt{5}s^2_\theta}{4} \\
\bottomrule  
\end{array}
\end{equation*}  
\end{small}

\subsubsection*{Coefficients appearing in $\mathcal{L}_{S V\tilde{V}}$}

The coefficients for the anomaly terms are specified up to an overall factor coming from the dimension of the hyperfermion irrep.
\begin{small}
\begin{equation*}
\def\arraystretch{1.4}
\begin{array}{cccccccccc}
\toprule  
& \tilde{K}^{S^{0}_i}_{\gamma\gamma} & \tilde{K}^{S^{0}_i}_{\gamma Z} & \tilde{K}^{S^{0}_i}_{ZZ} & \tilde{K}^{S^{0}_i}_{W^+W^-} & \tilde{K}^{S^{+}_i}_{\gamma W^-} & \tilde{K}^{S^{+}_i}_{ZW^-} & \tilde{K}^{S^{++}_i}_{W^-W^-}\\
\midrule 
h & 0 & \!\!\! 0 & \!\!\! 0 & \!\!\! 0 & \multicolumn{2}{c}{\multirow{5}{*}{$-$}} & \multirow{7}{*}{$-$} \\
\eta^0_3 & 0 & \!\!\! 0 & \!\!\! 0 & \!\!\! 0 \\
\eta^0_5 & -\frac{2s_\theta}{\sqrt{3}} & \!\!\! -\frac{c_{2W}s_\theta}{\sqrt{3}} & \!\!\! \frac{(1-3c_{4W}+2c_{2\theta})s_\theta}{12\sqrt{3}} & \!\!\! \frac{s^3_{\theta}}{6\sqrt{3}} \\
\eta^0_1 & \sqrt{\frac{2}{3}}s_\theta & \!\!\! \frac{c_{2W}s_\theta}{\sqrt{6}} & \!\!\! \frac{(1+6c_{4W}-7c_{2\theta})s_\theta}{24\sqrt{6}} & \!\!\! \frac{7s^3_{\theta}}{12\sqrt{6}} \\
\eta & \sqrt{\frac{2}{5}}s_\theta  & \!\!\! \frac{c_{2W}s_\theta}{\sqrt{10}} & \!\!\! \frac{(3c_\theta^2+c_{4W})s_\theta}{4\sqrt{10}} & \!\!\! \frac{(3c_{2\theta}+5)s_\theta}{8\sqrt{10}}\\
\eta^+_3 & \multicolumn{4}{c}{\multirow{2}{*}{$-$}} & -\frac{s_{2\theta}}{4} & \!\!\! \frac{s_W^2s_{2\theta}}{4} \\
\eta^+_5 & & & & & \frac{is_{\theta}}{2} & \!\!\! \frac{i(c_{2W}-c_{2\theta}-2)s_\theta}{12} \\
\eta^{++}_5 & \multicolumn{6}{c}{-} & -\frac{s_\theta^3}{3\sqrt{2}} \\
\bottomrule  
\end{array}
\end{equation*}  
\end{small}

\bibliographystyle{JHEP}
\bibliography{literature}

\end{document}